\documentclass[12pt]{article}
\usepackage{jheppub_modified_green}
\usepackage{epsfig}
\usepackage{amsfonts}
\usepackage{amssymb}
\usepackage{enumitem}
\usepackage{amsthm}
\usepackage{subfig}
\usepackage{bm}
\usepackage{hyperref}
\usepackage{mathrsfs}
\usepackage{bbm}
\usepackage{cancel}
\usepackage{xcolor}
\usepackage{accents}
\usepackage{relsize}
\usepackage{float, slashed, graphicx, amssymb, amsmath}
\usepackage{shuffle}
\usepackage{tikz}
\usetikzlibrary{hobby,calc,shapes, positioning, backgrounds,trees, decorations, decorations.markings, decorations.pathmorphing, arrows, shapes.geometric, snakes}
\usepackage{tikz-feynman}
\tikzfeynmanset{compat=1.1.0}
\tikzfeynmanset{
graviton/.style={circle, draw=green!60, fill=green!5, very thick, minimum size=7mm}
}
\tikzfeynmanset{myblob/.style=
{/tikz/shape=rectangle,
/tikz/fill=red,
 /tikz/minimum size=0.3cm,
 } }
\tikzset{box/.pic={\filldraw[fill=black]  (0,0) circle (2.5pt);
				   \filldraw [fill=black] (0.5,0) circle (2.5pt);
			       \draw [line width=5pt] (0,0) -- (0.5,0);}}

\usepackage[T1]{fontenc} 

\makeatletter
\newcommand \UPlus {\mathop {\operator@font \uplus }\limits }
\makeatother
\makeatletter
\newcommand \Bigcup {\mathop {\operator@font \bigcup }\limits }
\makeatother
  \def\LabelNote#1{}
 \def\Label#1{\label{#1}%
  \smash{\hbox to0pt{\raise1ex\hbox{\tiny[#1]}\hss}}}
  
  \def\Cdot{{\cdot}}
  
\def\nn{\nonumber}

\newcommand{\black}{\color{black}}

\def\spa#1.#2{\left\langle#1\,#2\right\rangle}
\def\spb#1.#2{\left[#1\,#2\right]}

\def\be{\begin{equation}}
\def\ee{\end{equation}}
\def\bea{\begin{eqnarray}}
\def\eea{\end{eqnarray}}  
\allowdisplaybreaks

\newcommand{\neco}{\mathbb{L}}          
\newcommand{\npre}{\mathcal{N}}  

\newcommand{\commut}{\Gamma}

\title{A new gauge-invariant double copy for   heavy-mass effective    theory}
\author{Andreas Brandhuber,}
\author{Gang Chen,}
\author{Gabriele Travaglini}
\author{and Congkao Wen}
\affiliation{Centre for Research in String Theory, School of Physics and Astronomy, \\
Queen Mary University of London, Mile End Road, London E1 4NS, United Kingdom}
\emailAdd{a.brandhuber@qmul.ac.uk}
\emailAdd{g.chen@qmul.ac.uk}
\emailAdd{g.travaglini@qmul.ac.uk}
\emailAdd{c.wen@qmul.ac.uk}

\begin{document} 
\begin{flushright}
	QMUL-PH-21-12\\
	SAGEX-21-04\\
\end{flushright}


\abstract{We propose a new form of the colour-kinematics/double-copy duality  for  heavy-mass effective field theories, which we apply to  construct compact expressions for tree amplitudes  with heavy matter particles in Yang-Mills and in gravity to leading order in the mass.
In this set-up,  the new BCJ numerators are fixed uniquely and directly written in terms of field strengths,  making  gauge invariance manifest. Furthermore, they are local  and automatically satisfy the Jacobi relations and crossing symmetry. We construct these  BCJ numerators explicitly up to six particles. We also discuss relations of the BCJ numerators  in the heavy-mass effective theory with those in pure Yang-Mills amplitudes. 

 }


\maketitle
\flushbottom

\newpage

\section{Introduction}

Problems in physics where the typical scale of the momenta is much smaller than the masses of the particles at play can be conveniently described using effective field theories. The prototype example is heavy-quark effective theory \cite{Georgi:1990um,Luke:1992cs,Neubert:1993mb,Manohar:2000dt}, where one is interested in studying the dynamics of heavy quarks exchanging momenta which are typically much smaller than their mass. The effective description is particularly convenient in that it reveals symmetries that are not present in the QCD Lagrangian, leading to a velocity superselection rule. Another example is the dynamics of black holes, which in many situations of practical relevance can be effectively considered  as heavy pointlike particles exchanging momenta which are much smaller than their  mass. 
In this context, the computation of  the classical part of observable quantities, such as the deflection angle of a massless particle by a heavy black hole, is of particular interest.  The classical limit is then reached by scaling the  momentum of the exchanged massless gravitons as $\vec{q} = \hbar \vec{k}$ and taking the $\hbar\to 0$ limit while keeping   the wavevector $\vec{k}$ fixed, together with the masses and momenta of the black holes \cite{Kosower:2018adc}. 
An approach to the effective  field theory describing heavy scalars and fermions minimally coupled to gravity in the spirit of the heavy-quark effective theory 
 was pursued in \cite{Damgaard:2019lfh}, with applications to Feynman-diagram computations of four-point amplitudes at one loop.

Recently, considerable effort has been devoted to applying modern amplitude methods to the computation of observable quantities in general relativity. This approach leads to significant conceptual as well as practical simplifications, since all steps of the computation are manifestly gauge invariant. Examples of this  include applications of unitarity at one  \cite{Neill:2013wsa,Bjerrum-Bohr:2013bxa,Bjerrum-Bohr:2014zsa, Bjerrum-Bohr:2016hpa, Cachazo:2017jef, Guevara:2017csg, Chi:2019owc,Brandhuber:2019qpg,Emond:2019crr,AccettulliHuber:2020oou}, two  \cite{Bern:2019nnu,Bern:2019crd,Cheung:2020gyp} and three loops \cite{Bern:2021dqo} to the computation of bending angles, and classical and quantum corrections to the gravitational potential.
An important tool in the amplitude arsenal is the  colour-kinematics and double-copy duality \cite{Bern:2008qj,Bern:2010ue,Bern:2019prr}, which   makes  explicit an intriguing interplay between colour and kinematics  of 
 amplitudes in Yang-Mills and other theories \cite{Bargheer:2012gv,Broedel:2012rc,Chiodaroli:2013upa,Johansson:2014zca,Chiodaroli:2014xia,Chiodaroli:2015rdg,Johansson:2017srf,Chiodaroli:2018dbu,Chen:2013fya,Carrasco:2016ldy,Mafra:2016mcc,Carrasco:2016ygv, Cheung:2016prv, Cheung:2017ems,Cheung:2017yef}. This also  allows one to construct gravity amplitudes from Yang-Mills ones once the latter are expressed in a so-called BCJ form, in a similar spirit to the  KLT relations \cite{Kawai:1985xq}. 
Further applications of this duality include  \cite{BjerrumBohr:2010hn,Mafra:2011kj,Fu:2012uy,Mafra:2015vca, Bjerrum-Bohr:2016axv,Du:2017kpo,Chen:2017bug, Fu:2018hpu,Bridges:2019siz} and  \cite{Bern:2012uf,Bjerrum-Bohr:2013iza,Bern:2013yya,Nohle:2013bfa,Mogull:2015adi,He:2017spx,Johansson:2017bfl, Kalin:2018thp,Boels:2012ew,Yang:2016ear,Boels:2017ftb} for the construction of BCJ numerators at tree and loop level, respectively.

Intriguingly, there is strong evidence that underlying the colour-kinematics duality there must be a kinematic algebra obeyed by the BCJ numerators. This algebra was originally discovered in the self-dual sector in \cite{Monteiro:2011pc,BjerrumBohr:2012mg}, with recent important efforts to understand it beyond that sector in \cite{Chen:2019ywi,Chen:2021chy}. We also mention the work of 
\cite{Fu:2016plh,Cheung:2016prv,Cheung:2017yef, Reiterer:2019dys,Tolotti:2013caa,Mizera:2019blq,Borsten:2020zgj,Borsten:2020xbt,Borsten:2021hua,Ferrero:2020vww} where a Lagrangian or geometric  understanding of the duality was sought.
With a view of applying the colour-kinematics duality to the problem of black hole scattering, one is interested in  amplitudes containing two massive particles as well as gravitons, which in general can also be constructed via the double copy \cite{Johansson:2015oia, delaCruz:2015dpa,Luna:2017dtq, Brown:2018wss,Plefka:2018zwm, Johansson:2019dnu,Bautista:2019evw, Plefka:2019wyg}. The double copy was also directly applied to the  heavy-quark effective theory  in  \cite{Haddad:2020tvs} for particles of spin  $s\leq 1$, in particular constructing three- and four-point amplitudes with two heavy spinning particles and one and two gravitons. 

In this paper we systematically apply the method of \cite{Chen:2019ywi} to obtain compact expressions for  amplitudes with two massive scalars and an arbitrary number of gluons or gravitons via the double copy in a heavy-mass effective theory (HEFT) at leading order in an inverse mass expansion.
These  will  be used in \cite{companion2} to compute loop  amplitudes of two heavy scalars with graviton interactions, from which one can extract classical quantities such as the bending angle and corrections  to the Newtonian potential. 
Our  HEFT amplitudes will  enter the relevant unitarity cuts,    crucially simplifying  the  loop integrations because of their 
special features \cite{companion2}.

In the  approach of  \cite{Chen:2019ywi}, one introduces vector and tensor currents representing the generators of the kinematic algebra, with a fusion rule among them. This fusion rule was completely determined in \cite{Chen:2019ywi, Chen:2021chy} in Yang-Mills in the MHV and NMHV sectors for arbitrary multiplicity, with explicit examples up to eight particles.%
\footnote{In general dimensions, by MHV amplitude in pure Yang-Mills we mean one  whose numerator has the schematic form  $(\epsilon \Cdot \epsilon)\prod (\epsilon \Cdot p)$, whereas NMHV corresponds to 
$(\epsilon \Cdot \epsilon)^2 \prod (\epsilon \Cdot p)$, and so~on.}
In the present work we  apply these ideas to the HEFT, computing  
 general amplitudes   up to six points and in any number of dimensions.
Importantly, these are the necessary  ingredients to compute the two-to-two scattering amplitude of two heavy scalars up to three loops.

This new approach is very powerful in that it leads to BCJ numerators automatically satisfying the Jacobi relations and crossing symmetry, under the assumption that the fusion rule is associative.   In practice,  one starts by building a function called the   pre-numerator (which, despite its name, can have  denominators from propagators of the heavy particles), which is written as a product of many currents multiplied using the fusion rule. 
The BCJ numerators are then obtained by taking appropriate anti-symmetrisations of the external particles in the pre-numerator, thereby forming nested commutator structures associated to cubic graphs. With the assumption of associativity of the fusion rule, this operation produces  
 BCJ numerators that automatically satisfy colour-kinematics duality. We also stress that only a subset of all possible cubic graphs appears in our construction, namely those where the two massive particles connect via a single cubic vertex to the rest of the graph. Each cubic graph is in one-to-one correspondence with a nested commutator.

 An additional important feature of our work is that the BCJ numerators we obtain  for each cubic graph are  uniquely determined, manifestly gauge invariant (i.e.~written in terms of field strengths) and local with respect to the massless gluons or gravitons. This is to be contrasted with the situation in Yang-Mills amplitudes, where the BCJ numerators are in general neither  gauge invariant nor  unique. As a byproduct of our analysis, we also show how to derive BCJ numerators in pure Yang-Mills by taking appropriate limits of our BCJ numerators.

 The rest of the paper is organised as follows. In the next section and in Section~\ref{sec:3} we briefly review basic properties of heavy-quark effective theory and of the double copy, respectively. In Section~\ref{sec:4} we present the construction of amplitudes from the novel double copy. We  briefly review the approach based on fusion rules,  which we then discuss in the context  of our HEFT. In particular we discuss the construction of the pre-numerator from fusion rules and from an ansatz, in terms of which the BCJ numerators are expressed. In Section~\ref{sec:5} we systematically treat cases up to six particles. Importantly, we find unique, gauge-invariant BCJ numerators. In Section~\ref{sec:6} we briefly discuss how to obtain pure Yang-Mills numerators from HEFT numerators in a particular limit. We present our conclusions and an outlook on future work in Section~\ref{sec:7}.

\section{Elements of heavy-mass effective theory}
\label{sec:2}

Heavy-quark effective theory \cite{Georgi:1990um,Luke:1992cs,Neubert:1993mb,Manohar:2000dt} plays an important role in hadron physics.  In this set-up, 
the momentum of an incoming heavy quark is written as
\begin{align}
p^\mu=m v^\mu
\ , 
\end{align}
where $m$ is the heavy mass of the quark and $v^2=1$, which after an interaction with a soft particle becomes
\begin{align}
p^\mu=m v^\mu+k^\mu
\ . 
\end{align}
In QCD, the momentum  $k^\mu$ would be taken to be of order $\Lambda_{\rm QCD} \ll m$. We are ultimately  interested in applications to 
classical physics (discussed in the companion paper \cite{companion2}),  in which case it is convenient to think of the residual soft momentum as being rescaled as $k^\mu = \hbar \bar{k}^\mu$ \cite{Kosower:2018adc}, keeping $\bar k$ fixed as $\hbar \to 0$.
If $p$ is the momentum of an on-shell state, for example an outgoing heavy quark, we also get the constraint $v\Cdot k= -k^2/(2m)$, which implies that $v \Cdot k = 0$ in the
large-mass limit.
The leading terms of the effective Lagrangian are 
\begin{align}
	\mathcal{L}_{\rm eff}= -\frac{1}{4}(F^a_{\mu\nu})^2+ i \bar Q_v v \Cdot (\partial - i g A) \frac{1 + \slashed{v}}{2}  Q_v + \mathcal{O}(1/m) \ ,
\end{align}
where for external fermion states one also has  $\slashed{v} Q_v=Q_v$. If one ignores the $\mathcal{O}(1/m)$ terms, the velocity $v$ and the polarisation of $Q_v$ are conserved. 
The Feynman rules for the fermion propagator and vertex are easily found to be 
\begin{align}
\label{feynmanrules}
\begin{tikzpicture}[baseline={([yshift=-0.8ex]current bounding box.center)}]\tikzstyle{every node}=[font=\small]
\begin{feynman}
\vertex (a);
\vertex  [right=2.0cm of a](b);
    \diagram*{
    (a) -- [fermion,thick,edge label'= {$v,k$}] (b),
    };
    \end{feynman}
  \end{tikzpicture}~~~
&&	{i\over v\Cdot k + i \varepsilon}{1+\slashed{v}\over 2}\, , \nn\\
\begin{tikzpicture}[baseline={([yshift=-0.8ex]current bounding box.center)}]\tikzstyle{every node}=[font=\small]    
    \begin{feynman}
    \vertex (a) {$p_1$};
    \vertex [right=1.0cm of a](f1);
    \vertex [right=0.8cm of f1](b){$p_3$};
    \vertex [above=0.8cm of f1](g2){$\mu, p_{2}$};
    	 \diagram*{(a) -- [fermion,thick,edge label'=$~~~~~~v$] (f1),(f1) -- [fermion,thick] (b),(f1)--[gluon](g2)};
    \end{feynman}  
  \end{tikzpicture}
  && i g T^a v_{\mu}  \frac{1 + \slashed{v}}{2}  \, , 
\end{align}
which are accompanied by the standard Feynman rules for gauge fields. 
Importantly, the leading-order contribution in the heavy-mass limit is universal, that is  the heavy quark field can be replaced by a heavy scalar or vector field without changing the amplitudes.

One can now use  these Feynman rules to compute directly HEFT amplitudes,  at least for a small number of legs.  For higher multiplicities this becomes very involved, and we will introduce more efficient techniques in the next two sections.%
\footnote{In the following we quote colour-ordered amplitudes and
drop an ubiquitous factor of~$ i \,  g^{n-2}$.}
The three-point amplitude
is given by
\begin{align}\label{eq:threeAmp}
	A^{\rm YM-M}_{3}(123)=\begin{tikzpicture}[baseline={([yshift=-0.8ex]current bounding box.center)}]\tikzstyle{every node}=[font=\small]	
\begin{feynman}
    	 \vertex (a) {\(p_1\)};
    	 \vertex [right=1.0cm of a] (f2);
    	 \vertex [right=0.8cm of f2] (c){$p_3$};
    	 \vertex [above=0.8cm of f2] (g2){$\varepsilon_2$};
    	  \diagram* {
(a) -- [fermion,thick] (f2)--  [fermion,thick] (c),
    	  (f2)--[gluon](g2)
    	  };
    \end{feynman}  
    \end{tikzpicture}= m \varepsilon_2\Cdot v  \ ,
\end{align}
while the four-point amplitude is 
\begin{align}
	A^{\rm YM-M}_{4}(1234)&=\begin{tikzpicture}[baseline={([yshift=-0.8ex]current bounding box.center)}]\tikzstyle{every node}=[font=\small]	
\begin{feynman}
    	 \vertex (a) {\(p_1\)};
    	 \vertex [right=0.8cm of a] (f2);
    	 \vertex [right=0.6cm of f2] (f3);
    	 \vertex [right=0.6cm of f3] (c){$p_4$};
    	 \vertex [above=0.8cm of f2] (g2){$p_2$};
    	 \vertex [above=0.8cm of f3] (g3){$p_3$};
    	  \diagram* {
(a) -- [fermion,thick] (f2)-- [fermion,thick] (f3) --  [fermion,thick] (c),
    	  (f2)--[gluon](g2), (f3)--[gluon](g3),
    	  };
    \end{feynman}  
    \end{tikzpicture}+\begin{tikzpicture}[baseline={([yshift=-0.8ex]current bounding box.center)}]\tikzstyle{every node}=[font=\small]	
\begin{feynman}
    	 \vertex (a) {\(p_1\)};
    	 \vertex [right=1.0cm of a] (f2);
    	 \vertex [right=0.8cm of f2] (c){$p_4$};
    	 \vertex [above=0.6cm of f2] (f3);
    	 \vertex [above left=0.8cm of f3] (g2){$p_2$};
    	 \vertex [above right=0.8cm of f3] (g3){$p_3$};
    	  \diagram* {
(a) -- [fermion,thick] (f2)-- [fermion,thick] (c), (f2)--[gluon](f3),
    	  (f3)--[gluon](g2), (f3)--[gluon](g3),
    	  };
    \end{feynman}  
    \end{tikzpicture}\nn\\
    &=2 m\Big(-\frac{\varepsilon _2\Cdot p_3 v\Cdot \varepsilon _3}{s_{23}}-\frac{\varepsilon _2\Cdot \varepsilon _3 v\Cdot p_2}{s_{23}}+\frac{\varepsilon _3\Cdot p_2 v\Cdot \varepsilon _2}{s_{23}}+\frac{v\Cdot \varepsilon _2 v\Cdot \varepsilon _3}{2 v\Cdot p_2}\Big) \ .
\end{align}
 The Feynman diagrams contributing to the five-point amplitude are 
 \begin{align}
	&\begin{tikzpicture}[baseline={([yshift=-0.8ex]current bounding box.center)}]\tikzstyle{every node}=[font=\small]
 \begin{feynman}
    	 \vertex (a) {\(p_1\)};
    	 \vertex [right=0.8cm of a] (f2);
    	 \vertex [right=0.8cm of f2] (f3);
    	 \vertex [right=0.8cm of f3] (f4);
    	 \vertex [right=0.6cm of f4] (c){$p_5$};
    	 \vertex [above=1cm of f2] (g2){$p_2$};
    	 \vertex [above=1cm of f3] (g3){$p_3$};
    	 \vertex [above=1cm of f4] (g4){$p_4$};
    	  \diagram* {
(a) -- [fermion,thick] (f2)-- [fermion,thick] (f3) -- [fermion,thick] (f4)-- [fermion,thick] (c),
    	  (f2)--[gluon](g2), (f3)--[gluon](g3),(f4)--[gluon](g4)
    	  };
    \end{feynman}  
  \end{tikzpicture}+\begin{tikzpicture}[baseline={([yshift=-0.8ex]current bounding box.center)}]\tikzstyle{every node}=[font=\small]
 \begin{feynman}
    	 \vertex (a) {\(p_1\)};
    	 \vertex [right=1.0cm of a] (f23);
    	 \vertex [right=1.2cm of f23] (f4);
    	 \vertex [right=0.8cm of f4] (c){$p_5$};
    	 \vertex [above=0.6cm of f23] (g23);
    	  \vertex [above left=0.6cm of g23] (g2){$p_2$};
    	 \vertex [above right=0.6cm of g23] (g3){$p_3$};
    	 \vertex [above=1.0cm of f4] (g4){$p_4$};
    	  \diagram* {
(a) -- [fermion,thick] (f23) -- [fermion,thick] (f4)-- [fermion,thick] (c),
    	  (f23)--[gluon](g23), (g23)--[gluon](g3),(g23)--[gluon](g2),(f4)--[gluon](g4)
    	  };
    \end{feynman}  
  \end{tikzpicture}+ \begin{tikzpicture}[baseline={([yshift=-0.8ex]current bounding box.center)}]\tikzstyle{every node}=[font=\small]
  \begin{feynman}
    	 \vertex (a) {\(p_1\)};
    	 \vertex [right=1.0cm of a] (f2);
    	 \vertex [right=1.2cm of f2] (f34);
    	 \vertex [right=0.8cm of f34] (c){$p_5$};
    	 \vertex [above=0.6cm of f34] (g34);
    	 \vertex [above=1.0cm of f2] (g2){$p_2$};
    	  \vertex [above left=0.6cm of g34] (g3){$p_3$};
    	 \vertex [above right=0.6cm of g34] (g4){$p_4$};
    	  \diagram* {
(a) -- [fermion,thick] (f2) -- [fermion,thick] (f34)-- [fermion,thick] (c),(f2)--[gluon](g2), (g34)--[gluon](g3),(g34)--[gluon](g4),(f34)--[gluon](g34)
    	  };
    \end{feynman}  
  \end{tikzpicture}\nn\\
 +& \begin{tikzpicture}[baseline={([yshift=-0.8ex]current bounding box.center)}]\tikzstyle{every node}=[font=\small]
    \begin{feynman}
    	 \vertex (a) {\(p_1\)};
    	 \vertex [right=of a] (b);
    	  \vertex [right=of b] (c){$p_5$};
    	  \vertex [above=0.8cm of b] (g234);
    	  \vertex [above right=1.6cm of g234] (g4){$p_4$};
    	  \vertex [above left=0.8cm of g234] (g23);
    	   \vertex [above right=0.8cm of g23] (g3){$p_3$};
    	   \vertex [above left=0.8cm of g23] (g2){$p_2$};
    	  \diagram* {
(a) -- [fermion,thick] (b) -- [fermion,thick] (c),
    	  (b)--[gluon](g234), (g234)--[gluon](g4),(g234)--[gluon](g23),(g23)--[gluon](g3),(g23)--[gluon](g2)
    	  };
    \end{feynman}  
  \end{tikzpicture}+\begin{tikzpicture}[baseline={([yshift=-0.8ex]current bounding box.center)}]\tikzstyle{every node}=[font=\small]
    \begin{feynman}
    	 \vertex (a) {\(p_1\)};
    	 \vertex [right=of a] (b);
    	  \vertex [right=of b] (c){$p_5$};
    	  \vertex [above=0.8cm of b] (g234);
    	  \vertex [above left=1.6cm of g234] (g2){$p_2$};
    	  \vertex [above right=0.8cm of g234] (g34);
    	   \vertex [above right=0.8cm of g34] (g4){$p_4$};
    	   \vertex [above left=0.8cm of g34] (g3){$p_3$};
    	  \diagram* {
(a) -- [fermion,thick] (b) -- [fermion,thick] (c),
    	  (b)--[gluon](g234), (g234)--[gluon](g2),(g234)--[gluon](g34),(g34)--[gluon](g3),(g34)--[gluon](g4)
    	  };
    \end{feynman}  
  \end{tikzpicture}+\begin{tikzpicture}[baseline={([yshift=-0.8ex]current bounding box.center)}]\tikzstyle{every node}=[font=\small]
    \begin{feynman}
    	 \vertex (a) {\(p_1\)};
    	 \vertex [right=of a] (b);
    	  \vertex [right=of b] (c){$p_5$};
    	  \vertex [above=0.8cm of b] (g234);
    	  \vertex [above left=1.6cm of g234] (g2){$p_2$};
    	  \vertex [above right=0.8cm of g234] (g34);
    	   \vertex [above right=0.8cm of g34] (g4){$p_4$};
    	   \vertex [right=1.5cm of g2] (g3){$p_3$};
    	  \diagram* {
(a) -- [fermion,thick] (b) -- [fermion,thick] (c),
    	  (b)--[gluon](g234), (g234)--[gluon](g2),(g234)--[gluon](g3),(g234)--[gluon](g4)
    	  };
    \end{feynman}  
  \end{tikzpicture}\nn 
  \end{align}
leading to the -- somewhat lengthy -- result
  \begin{align}
  \label{fivepoints}
 	A^{\rm YM-M}_{5}(12345) =  & \ \ 4 m\bigg[\frac{v\Cdot \varepsilon _2 v\Cdot \varepsilon _3 v\Cdot \varepsilon _4}{4 v\Cdot p_2 v\Cdot p_{23}}+\frac{v\Cdot \varepsilon _4 \left(\frac{1}{2} \varepsilon _2\Cdot \varepsilon _3 v\Cdot (p_3-p_2)+\varepsilon _3\Cdot p_2 v\Cdot \varepsilon _2-\varepsilon _2\Cdot p_3 v\Cdot \varepsilon _3\right)}{2 s_{23} v\Cdot p_{23}}\nn\\
  +&\frac{v\Cdot \varepsilon _2 \left(-\varepsilon _3\Cdot p_4 v\Cdot \varepsilon _4-\frac{1}{2} \varepsilon _3\Cdot \varepsilon _4 v\Cdot p_3+\frac{1}{2} \varepsilon _3\Cdot \varepsilon _4 v\Cdot p_4+\varepsilon _4\Cdot p_3 v\Cdot \varepsilon _3\right)}{2 s_{34} v\Cdot p_2}\nn\\
  +&\frac{1}{2 s_{23} s_{234}} \Big(v\Cdot \varepsilon _4 \left(({s_{24}-s_{34}\over 2}) \varepsilon _2\Cdot \varepsilon _3-2 \varepsilon _2\Cdot p_4 \varepsilon _3\Cdot p_2+2 \varepsilon _2\Cdot p_3 \varepsilon _3\Cdot p_4\right)\nn\\
  +&{s_{23}\over 2} \left(\varepsilon _2\Cdot \varepsilon _4 v\Cdot \varepsilon _3-\varepsilon _3\Cdot \varepsilon _4 v\Cdot \varepsilon _2\right)-\varepsilon _4\Cdot p_{23} \left(\varepsilon _2\Cdot p_3 v\Cdot \varepsilon _3-\varepsilon _3\Cdot p_2 v\Cdot \varepsilon _2\right)\nn\\
  +&2 \left( v\Cdot p_{23}\varepsilon _3\Cdot \varepsilon _4 \varepsilon _2\Cdot p_3-v\Cdot p_{23}\varepsilon _2\Cdot \varepsilon _4 \varepsilon _3\Cdot p_2+v\Cdot p_3\varepsilon _2\Cdot \varepsilon _3 \varepsilon _4\Cdot p_2-v\Cdot p_2\varepsilon _2\Cdot \varepsilon _3 \varepsilon _4\Cdot p_3\right)\Big)\nn\\
  +&\frac{1}{s_{34} s_{234}}\Big(\varepsilon _2\Cdot p_{34} (\varepsilon _3\Cdot p_4 v\Cdot \varepsilon _4- \varepsilon _4\Cdot p_3 v\Cdot \varepsilon _3)+\varepsilon _2\Cdot \varepsilon _4 v\Cdot p_2 \varepsilon _3\Cdot p_4-\varepsilon _4\Cdot p_2 \varepsilon _3\Cdot p_4 v\Cdot \varepsilon _2\nn\\
  +&\varepsilon _3\Cdot \varepsilon _4 v\Cdot p_3 \varepsilon _2\Cdot p_{34}+\varepsilon _3\Cdot \varepsilon _4 v\Cdot p_2 \varepsilon _2\Cdot p_3-\varepsilon _2\Cdot \varepsilon _3 v\Cdot p_2 \varepsilon _4\Cdot p_3+\varepsilon _3\Cdot p_2 \varepsilon _4\Cdot p_3 v\Cdot \varepsilon _2\nn\\
  -&\frac{1}{4} s_{23} \varepsilon _3\Cdot \varepsilon _4 v\Cdot \varepsilon _2+\frac{1}{4} s_{24} \varepsilon _3\Cdot \varepsilon _4 v\Cdot \varepsilon _2-\frac{1}{4} s_{34} \varepsilon _2\Cdot \varepsilon _3 v\Cdot \varepsilon _4+\frac{1}{4} s_{34} \varepsilon _2\Cdot \varepsilon _4 v\Cdot \varepsilon _3\Big) \bigg],
\end{align}
where  
\begin{align}
p_{i_1i_2\cdots i_r}:= p_{i_1}+p_{i_2}+\cdots+p_{i_r}\, , \qquad s_{i_1i_2\cdots i_r}:=p_{i_1i_2\cdots i_r}^2\ . 
\end{align}
Note that the gluon-quark amplitude in HEFT does not depend on the soft momentum of the heavy particle. One can also check that the leading order term in $1/m$ of the full five-point gluon-quark amplitude is identical to \eqref{fivepoints}.


\section{Traditional double copy construction}
\label{sec:3}

In the previous section we  reviewed an example of a heavy-particle effective theory and presented 
a few tree-level amplitudes derived from the leading-order Feynman rules. We chose to work with fermions as heavy particles, 
however it is important to note that the results thus obtained are in fact independent of the spin of the heavy matter fields  \cite{Bjerrum-Bohr:2013bxa}. The corresponding effective theory for gravity, studied using  a form of the colour-kinematics duality and double copy of HEFT in \cite{Haddad:2020tvs}, also exhibits universal behaviour at leading order in the heavy-mass limit, which is related to the universal coupling between matter and gravitons \cite{Heisenberg:2018vsk,Minazzoli:2013bva}.

Next, we discuss  quark-graviton amplitudes for the full theory as  computed using the traditional form of the colour-kinematics duality. The leading-order parts of these amplitudes in the heavy-mass limit can  then be obtained by an expansion in the (inverse) heavy mass.%
\footnote{The quark-graviton amplitudes can also be obtained from  Feynman rules \cite{huggins1962quantum,sannan1986gravity}.}
In this form of the  double copy one sums over \emph{all} cubic graphs,  and  the numerators  are in general \emph{not} gauge invariant.
The double copy of QCD with massive fermions was studied in \cite{Johansson:2019dnu,Haddad:2020tvs}, while the double copy of massive scalar QCD was investigated in \cite{Plefka:2019wyg}.

The BCJ numerators for the three-, four- and five-point amplitudes can be cast in the form \cite{Johansson:2014zca,Johansson:2019dnu,Chen:2019ywi}
\begin{align}
\label{N345}
	N_3(123)&=\bar Q \slashed\varepsilon_2 Q,\nn\\
	N_4(1234)&= \bar Q\slashed\varepsilon _2 (\slashed p_{12}+m)\slashed \varepsilon _3 Q,\nn\\
	N_5(12345)&=\bar Q \slashed \varepsilon _2(\slashed p_{12}+m) \slashed\varepsilon _3 (\slashed p_{123}+m)\slashed \varepsilon_4 Q+\frac{2}{3}  (p_{123}^2-m^2) \left(\bar Q \slashed\varepsilon _2 \slashed \varepsilon _3\slashed \varepsilon _4 Q-\varepsilon _3\cdot \varepsilon _4\bar Q \slashed\varepsilon _2 Q \right) \ .
\end{align}
The three-point graviton amplitude is obtained by just squaring  the BCJ numerator
\begin{align}\label{eq:A3GR}
	A^{\rm GR-Q}_{3}&=\big[N_3(123)\big]^2\ .
\end{align}
The  four-point amplitude is obtained by summing over the following three  cubic graphs with two gravitons and two fermions: 
\begin{align}
& \begin{tikzpicture}[baseline={([yshift=-0.8ex]current bounding box.center)}]\tikzstyle{every node}=[font=\small]	
\begin{feynman}
    	 \vertex (a) {$p_1$};
    	 \vertex [right=1.0cm of a] (f2);
    	 \vertex [right=0.6cm of f2] (f3);
    	 \vertex [right=0.6cm of f3] (c){$p_4$};
    	 \vertex [above=0.8cm of f2] (g2){$p_2$};
    	 \vertex [above=0.8cm of f3] (g3){$p_3$};
    	  \diagram* {
(a) -- [fermion,thick] (f2)-- [fermion,thick] (f3) --  [fermion,thick] (c),
    	  (f2)--[photon,ultra thick](g2), (f3)--[photon,ultra thick](g3)
    	  };
    \end{feynman}  
    \end{tikzpicture}+\begin{tikzpicture}[baseline={([yshift=-0.8ex]current bounding box.center)}]\tikzstyle{every node}=[font=\small]	
\begin{feynman}
    	 \vertex (a) {$p_1$};
    	 \vertex [right=1.0cm of a] (f2);
    	 \vertex [right=0.8cm of f2] (f3);
    	 \vertex [right=0.6cm of f3] (c){$p_4$};
    	 \vertex [above right=1.0cm of f2] (g3){$p_3$};
    	 \vertex [above left=1.0cm of f3] (g2){$p_2$};
    	  \diagram*{
(a) -- [fermion,thick] (f2)-- [fermion,thick] (f3) --  [fermion,thick] (c),
    	  (f2)--[photon,ultra thick](g3), (f3)--[photon,ultra thick](g2)
    	  };
    \end{feynman}  
    \end{tikzpicture} +
    \begin{tikzpicture}[baseline={([yshift=-0.8ex]current bounding box.center)}]\tikzstyle{every node}=[font=\small]	
\begin{feynman}
    	 \vertex (a) {$p_1$};
    	 \vertex [right=1.0cm of a] (f2);
    	 \vertex [right=0.6cm of f2] (c){$p_4$};
    	 \vertex [above=0.5cm of f2] (f3);
    	 \vertex [above left=0.8cm of f3] (g2){$p_2$};
    	 \vertex [above right=0.8cm of f3] (g3){$p_3$};
    	  \diagram* {
(a) -- [fermion,thick] (f2)-- [fermion,thick] (c), (f2)--[photon,ultra thick](f3),
    	  (f3)--[photon,ultra thick](g2), (f3)--[photon,ultra thick](g3)
    	  };
    \end{feynman}  
    \end{tikzpicture}.
\end{align}
The numerator of  each graph is the square of the corresponding BCJ numerator.  The four-point amplitude is then
\begin{align}\label{eq:A4GR}
	A_{4}^{\rm GR-Q}&={N_4(1234)^2\over 2p_1\Cdot p_2}+{N_4(1324)^2\over 2p_1\Cdot p_3}+{N_4(1[2,3]4)^2\over s_{23}}\ ,
\end{align}
where $N_4(1324)$ is obtained from $N_4(1234)$ by  swapping the indices $2,3$ and the bracket on the external labels denotes the  commutator of the indices, e.g. 
\begin{align}N_4(1[2,3]4):=N_4(1234)-N_4(1324)
\ .
\end{align}
The four-point amplitude is manifestly invariant under the exchange of particles  $2$ and $3$. Similarly the five-point amplitude is obtained by summing  over  all cubic graphs with two scalars and three external gravitons:
\begin{align}
&\begin{tikzpicture}[baseline={([yshift=-0.8ex]current bounding box.center)}]\tikzstyle{every node}=[font=\small]
 \begin{feynman}
    	 \vertex (a) {\(p_1\)};
    	 \vertex [right=0.8cm of a] (f2);
    	 \vertex [right=0.8cm of f2] (f3);
    	 \vertex [right=0.8cm of f3] (f4);
    	 \vertex [right=0.6cm of f4] (c){$p_5$};
    	 \vertex [above=1cm of f2] (g2){$p_2$};
    	 \vertex [above=1cm of f3] (g3){$p_3$};
    	 \vertex [above=1cm of f4] (g4){$p_4$};
    	  \diagram* {
(a) -- [fermion,thick] (f2)-- [fermion,thick] (f3) -- [fermion,thick] (f4)-- [fermion,thick] (c),
    	  (f2)--[photon,ultra thick](g2), (f3)--[photon,ultra thick](g3),(f4)--[photon,ultra thick](g4)
    	  };
    \end{feynman}  
  \end{tikzpicture}+\begin{tikzpicture}[baseline={([yshift=-0.8ex]current bounding box.center)}]\tikzstyle{every node}=[font=\small]
 \begin{feynman}
    	 \vertex (a) {\(p_1\)};
    	 \vertex [right=1.0cm of a] (f23);
    	 \vertex [right=1.2cm of f23] (f4);
    	 \vertex [right=0.8cm of f4] (c){$p_5$};
    	 \vertex [above=0.6cm of f23] (g23);
    	  \vertex [above left=0.6cm of g23] (g2){$p_2$};
    	 \vertex [above right=0.6cm of g23] (g3){$p_3$};
    	 \vertex [above=1.0cm of f4] (g4){$p_4$};
    	  \diagram* {
(a) -- [fermion,thick] (f23) -- [fermion,thick] (f4)-- [fermion,thick] (c),
    	  (f23)--[photon,ultra thick](g23), (g23)--[photon,ultra thick](g3),(g23)--[photon,ultra thick](g2),(f4)--[photon,ultra thick](g4)
    	  };
    \end{feynman}  
  \end{tikzpicture}+ \begin{tikzpicture}[baseline={([yshift=-0.8ex]current bounding box.center)}]\tikzstyle{every node}=[font=\small]
  \begin{feynman}
    	 \vertex (a) {\(p_1\)};
    	 \vertex [right=1.0cm of a] (f2);
    	 \vertex [right=1.2cm of f2] (f34);
    	 \vertex [right=0.8cm of f34] (c){$p_5$};
    	 \vertex [above=0.6cm of f34] (g34);
    	 \vertex [above=1.0cm of f2] (g2){$p_2$};
    	  \vertex [above left=0.6cm of g34] (g3){$p_3$};
    	 \vertex [above right=0.6cm of g34] (g4){$p_4$};
    	  \diagram* {
(a) -- [fermion,thick] (f2) -- [fermion,thick] (f34)-- [fermion,thick] (c),(f2)--[photon,ultra thick](g2), (g34)--[photon,ultra thick](g3),(g34)--[photon,ultra thick](g4),(f34)--[photon,ultra thick](g34)
    	  };
    \end{feynman}  
  \end{tikzpicture}\nn\\
 +& \begin{tikzpicture}[baseline={([yshift=-0.8ex]current bounding box.center)}]\tikzstyle{every node}=[font=\small]
    \begin{feynman}
    	 \vertex (a) {\(p_1\)};
    	 \vertex [right=of a] (b);
    	  \vertex [right=of b] (c){$p_5$};
    	  \vertex [above=0.3cm of b] (g234);
    	  \vertex [above right=1.6cm of g234] (g4){$p_4$};
    	  \vertex [above left=0.8cm of g234] (g23);
    	   \vertex [above right=0.8cm of g23] (g3){$p_3$};
    	   \vertex [above left=0.8cm of g23] (g2){$p_2$};
    	  \diagram* {
(a) -- [fermion,thick] (b) -- [fermion,thick] (c),
    	  (b)--[photon,ultra thick](g234), (g234)--[photon,ultra thick](g4),(g234)--[photon,ultra thick](g23),(g23)--[photon,ultra thick](g3),(g23)--[photon,ultra thick](g2)
    	  };
    \end{feynman}  
  \end{tikzpicture}+\begin{tikzpicture}[baseline={([yshift=-0.8ex]current bounding box.center)}]\tikzstyle{every node}=[font=\small]
    \begin{feynman}
    	 \vertex (a) {\(p_1\)};
    	 \vertex [right=of a] (b);
    	  \vertex [right=of b] (c){$p_5$};
    	  \vertex [above=0.3cm of b] (g234);
    	  \vertex [above left=1.6cm of g234] (g2){$p_2$};
    	  \vertex [above right=0.8cm of g234] (g34);
    	   \vertex [above right=0.8cm of g34] (g4){$p_4$};
    	   \vertex [above left=0.8cm of g34] (g3){$p_3$};
    	  \diagram* {
(a) -- [fermion,thick] (b) -- [fermion,thick] (c),
    	  (b)--[photon,ultra thick](g234), (g234)--[photon,ultra thick](g2),(g234)--[photon,ultra thick](g34),(g34)--[photon,ultra thick](g3),(g34)--[photon,ultra thick](g4)
    	  };
    \end{feynman}  
  \end{tikzpicture}+\text{permutations of 2,3,4.}
  \end{align}
  Computing the above diagrams, one gets
\begin{align}\label{eq:A5GR}
		A_{5}^{\rm GR-Q}&={N_5(12345)^2\over 2p_1\Cdot p_2 (2p_1\Cdot p_{23}+s_{23})}+{N_5(1[2,3]45)^2\over 2s_{23} (2p_1\Cdot p_{23}+s_{23})}+{N_5(12[3,4]5)^2\over 4s_{34} p_1\Cdot p_2}\nn\\
 	&+{N_5(1[2,[3,4]]5)^2\over 2s_{34} s_{234}}\ + \ \text{permutations of}\ 2,3,4 \ .
\end{align}
The leading-order term in the heavy-mass expansion is  obtained by simply retaining  the $\mathcal{O}(m^2)$ terms in~\eqref{eq:A5GR}, and we have checked that it   agrees with the same quantity as computed with the KLT duality.

A few comments are in order here. 
First, we note that in this traditional form of the double copy one has to sum over all possible cubic graphs; furthermore,  gauge invariance
of the amplitudes
is not manifest due to the explicit appearance of polarisation vectors in the numerators $N_i$. 
Inspired by recent work on the kinematic algebra of BCJ numerators  \cite{Chen:2019ywi}, 
in the next section we will introduce an improved version of the double copy, which will lead  to much more compact expressions for the amplitudes. These will be  used in  \cite{companion2} for the  computation of classical quantities in gravity  at  loop level.

\section{A novel  double copy from gauge-invariant  BCJ  numerators}
\label{sec:4}

We now present our alternative form of the colour-kinematics duality  for the gluon-matter amplitudes directly  at  leading order in the HEFT, avoiding the need to perform a heavy-mass expansion.  
This double copy is motivated by the work of one of the present authors \cite{Chen:2019ywi} on the algebraic structure of 
numerators that are consistent with the colour-kinematics duality, and allows us to generate
the BCJ numerators directly. 
As we will see,  this construction has several advantages: 
\begin{itemize}
\item[{\bf 1.}] The new BCJ numerators  are automatically gauge invariant and unique. 
\item[{\bf 2.}] Only a subset of the usual cubic diagrams contributes.
\item[{\bf 3.}] As a consequence, we obtain  much more compact expressions for the amplitudes than those derived from the traditional BCJ construction discussed in the previous section. 
\end{itemize}
We have tested this new method by explicitly constructing numerators up to six particles at tree level, although in principle  the method   applies to $n$-particle numerators.

\subsection{Background for the colour-kinematics algebra from fusion rules}
We now describe the new construction of BCJ numerators, which is based on the {\it fusion product} between two heavy-mass currents.  Following the notation  of
 \cite{Chen:2019ywi} these currents are denoted by   $J_{a_1\otimes a_2\otimes\cdots\otimes a_r}$,%
 \footnote{The notation for  these currents is inspired   by the form of tensor currents in QCD, 
$
\bar Q \slashed a_1 \slashed a_2 \cdots \slashed a_r Q$.}
and their  labels $a_i^\mu$  can   either   be momenta or polarisation vectors. Since  $\bar Q \slashed a_i Q\sim m v\Cdot a_i$  in  the HEFT, we define
\begin{align}
\label{4.1}
J_{a}\, =\, m\,  v\Cdot a\ .
\end{align}
We also require that the  tensor currents $J_{a_1\otimes a_2\cdots a_i \otimes a_j\cdots\otimes a_r}$ satisfy the Clifford algebra 
 \begin{align}\label{eq:CliffordAl}
J_{a_1\otimes a_2\cdots a_i \otimes a_j\cdots\otimes a_r}=-J_{a_1\otimes a_2\cdots a_j \otimes a_i\cdots\otimes a_r} \, +\, 2 \, a_i\Cdot a_j J_{a_1\otimes a_2\cdots \hat a_i \otimes \hat a_j\cdots\otimes a_r}\ , 
\end{align} for each component. This property is  inherited from the  QCD currents. 

The on-shell condition for the external quarks also gives the relations
\begin{align}
\begin{split}
\label{eq:onshellJ}
J_{a_1\otimes a_2 \otimes \cdots a_n \otimes p_{12\ldots n-1}} & = m \, J_{a_1\otimes a_2 \otimes \cdots \otimes a_n } \ ,\\
J_{p_1 \otimes a_1\otimes a_2 \otimes \cdots \otimes a_n } & =  m  \, J_{a_1\otimes a_2 \otimes \cdots \otimes a_n} \ ,
\end{split}
\end{align}
and the currents satisfy a bilinear fusion rule  of the form 
\begin{align}
\label{fusionrule}
J_X\star J_Y =\sum_Z F_{XY}^{\ \ \ Z} J_Z\ . 
\end{align}
For instance \cite{Chen:2019ywi}
\begin{align}
\label{strconst4}
J_{\varepsilon_2} \star J_{\varepsilon_3} := 2\Big(
\frac{s_{23}}{4} \frac{v\Cdot \varepsilon _3}{v\Cdot p_2}J_{ \varepsilon _2} -{1\over 2} J_{\varepsilon _2\otimes \varepsilon _3\otimes p_2}+\varepsilon _3\Cdot p_2 J_{ \varepsilon _2}\Big)
\ . 
\end{align}
The fusion rule always generates a   rational function which we call the  {\it pre-numerator}: \begin{align}
\npre_{n}(23\cdots n-1, v):=
\begin{tikzpicture}[baseline={([yshift=-0.8ex]current bounding box.center)}]\tikzstyle{every node}=[font=\small]    
    \begin{feynman}
    \vertex (a)[myblob]{};
     \vertex [ left=1.2cm of a](b);
    \vertex [above=.8cm of b](j1){$2$};
    \vertex [right=1.2cm of j1](j2){$3$};
    \vertex [right=1.2cm of j2](j3){$n-1$};
    \vertex [right=0.5cm of j2](jnote){$\cdots$};
   	 \diagram*{(a) -- [thick] (j1),(a) -- [thick] (j2),(a)--[thick](j3)};
    \end{feynman}  
  \end{tikzpicture} 	  &=J_{\varepsilon_2}\star J_{\varepsilon_3} \star \cdots \star J_{\varepsilon_{n-1}},
\end{align} 
where we always assume  associativity of the $\star$-product.  In this diagram, the red box denotes the two massive particles, while the lines correspond to the massless particles (gluons or gravitons).
The general form of the fusion rule with these properties is known in pure Yang-Mills from  \cite{Chen:2019ywi} for $n$-point MHV amplitudes and for NMHV amplitudes up to eight points. The claim is that BCJ numerators in HEFT can be written efficiently in terms of the pre-numerator, in a way we describe below.

To begin with,  it is  useful to introduce the notion of  ordered and un-ordered nested commutators.
In the case of ordered nested commutators, which is relevant for colour-ordered amplitudes, the order of a set of indices is fixed while commutators are
applied in all possible ways. For example, for $n=5$, the ordered set $\{2,3,4\}$  gives rise to two ordered, nested commutators: $[[2,3],4]$ and $[2,[3,4]]$.
In the case of graviton amplitudes, we need to include also   un-ordered nested commutators (omitting numerators that differ by minus signs): 
$[[2,3],4]$, $[[2,4],3]$ and $[[3,4],2]$. 
Then the gluon-matter and graviton-matter amplitudes are given by the following expressions:
\begin{align}
\begin{split}
\label{eq:newDC}
	A^{\rm YM-M}_n(12\cdots n)&=\sum_{\commut \in \text{ordered commutators } \{2,3,\cdots,n-1\}} {\npre_n(\commut,v)\over d_\commut}\, ,
	\\
	A^{\rm GR-M}_n(12\cdots n)&=\sum_{\commut\in \text{non-ordered commutators} \{2,3,\cdots,n-1\}} {\big[\npre_n(\commut,v)\big]^2 \over d_\commut}\, ,
\end{split}
\end{align}
where  particles $1$ and $n$ are  heavy  and all others  are massless. 
Each nested commutator (and hence BCJ numerator) is in one-to-one correspondence with a specific cubic graph, from which one can also read off the relevant massless scalar propagators, denoted as $d_\commut$ in   \eqref{eq:newDC}.  For instance, the nested commutator $[[2,3],4]$ and the associated BCJ numerator    corresponds to the cubic graph  
\begin{align}
\npre_5([[2,3],4],v)\ \ \ \longleftrightarrow \ \ \ 
\begin{tikzpicture}[baseline={([yshift=-0.8ex]current bounding box.center)}]\tikzstyle{every node}=[font=\small]    
   \begin{feynman}
    \vertex (a)[myblob]{};
     \vertex [above=0.5cm of a](b)[dot]{};
     \vertex [left=1.0cm of b](c);
     \vertex [left=0.5cm of b](c23);
     \vertex [above=0.35cm of c23](v23)[dot]{};
    \vertex [above=.6cm of c](j1){$2$};
    \vertex [right=1.0cm of j1](j2){$3$};
    \vertex [right=1.0cm of j2](j3){$4$};
   	 \diagram*{(a) -- [thick] (b),(b) -- [thick] (j1),(v23) -- [thick] (j2),(b)--[thick](j3)};
    \end{feynman}  
  \end{tikzpicture}
\ , 
\end{align}
where in this case $d_{[[2,3],4]} = s_{23} s_{234}$.  We also note an important feature of this BCJ representation, namely that we only sum over  cubic graphs corresponding to nested commutators. In particular, only massless propagators appear within the cubic graph -- the two massive particles always connect to the graph via a single cubic vertex.

Each BCJ numerator $\npre_n(\commut,v)$ can be conveniently  obtained by acting with specific operators on the pre-numerator.  
For instance,  a left-nested commutator is  written as 
 \begin{align}
 	\npre_n([\ldots[[2,3],4],\ldots, n-1],v ):= \mathbb{L}(2,3,4,\ldots, n-1)\circ  \npre_n(234\ldots n-1,v), 
 \end{align}
where $\mathbb{L}(2,3,4,\ldots, n-1)$ is the  $\mathbb L$-operator  introduced  in \cite{Chen:2021chy} as a group algebra  element 
\cite{linckelmann2018block,algebra}
and  in \cite{reutenauer2003free} as a free Lie algebra element, 
\begin{align}
\neco(i_1, i_2,\ldots, i_r):= \Big[\mathbb{I}-\mathbb{P}_{(i_2i_1)}\Big]\Big[\mathbb{I}-\mathbb{P}_{(i_3i_2i_1)}\Big]\cdots \Big[ \mathbb{I}-\mathbb{P}_{(i_r\ldots i_2i_1)}\Big]\, .
\end{align}
Here  $\mathbb{P}_{(j_1 j_2 j_3 \ldots j_m)}$  denotes the cyclic permutation   $j_1\to j_2 \to j_3\to \cdots \to j_m\to j_1$. For instance 
\begin{align}
\begin{split}
	\mathbb I \circ\npre_6(2345, v)&= \npre_6(2345,v )\, , \\
	\black \mathbb P_{(432)} \circ \npre_6(2345,v)&=\npre_6(4235,v)\, .
\end{split}
\end{align}
Importantly, as we will discuss in more detail later, for each choice of $\commut$ the numerator $\npre_n(\commut,v)$ is  consistent with  colour-kinematics duality.

We have successfully checked this conjecture  up to six points. 
An important observation is  that the  numerators we   obtain in this approach  are  all  gauge invariant, as will become clear in concrete examples presented later.   
The general method to determine the fusion rules will be described elsewhere,  while in this paper we will focus on describing a  method to construct   the pre-numerator from an  ansatz, and from this, the BCJ numerators 
$\npre_n(\commut,v)$.

\subsection{General method to construct the pre-numerator}
\label{sec:Ansatz}

 Since the fusion rule introduced in \eqref{fusionrule} is not yet known in general, our strategy consists in 
 writing down a general  ansatz for the pre-numerator, in terms of  which 
   the BCJ numerators $\npre_n(\commut,v)$ are written. 
 We  then require that  these  BCJ numerators $\npre_n(\commut,v)$   generate the correct amplitudes 
when inserted  in  \eqref{eq:newDC}. 
 In  all cases we have considered,  this procedure leads to a unique, gauge-invariant answer for the BCJ numerators 
  (while in general the pre-numerator has undetermined coefficients). 
   
 The building blocks of   the ansatz are  products of the following Lorentz-invariant quantities:%
 \footnote{Although $J_{a}$ is proportional to  $v\Cdot a$, we prefer to treat these two quantities as separate as their origin in the QCD amplitude is different, specifically the $J_a$ terms arise from fermion bilinears while the $v\Cdot a$ terms from expanding propagators. 
Note that in constructing the ansatz,  we always fix $J_{a}$ to be either $J_{p_2}$ or  $J_{\varepsilon_2}$ from the requirement of full crossing symmetry with respect to the  gravitons.} 
 \begin{align}
 	{1\over v\Cdot p_{i_1\ldots i_r}}\, , \quad  \varepsilon_i\Cdot \varepsilon_j\, ,\quad  \varepsilon_i\Cdot p_j \, \quad v\Cdot\varepsilon_i \, , \quad v\Cdot p_i\, , \quad s_{i j}  \,  , \quad J_{\varepsilon_i}\, , \quad  J_{p_i}\, .
 \end{align}
In fact,  in  constructing  the ansatz we can restrict the vector currents $J$ to only those involving $J_{p_2}$ and $J_{\varepsilon_2}$, since the  others   can be generated with the $\mathbb L(2,3,4,\ldots, n-1)$ operator.

There are several constraints to impose on  our ansatz. First, for an $n$-point Yang-Mills amplitude, there are at most $n-3$ propagators of the heavy particles and no double poles.  Second,  it follows from the Feynman rules \eqref{feynmanrules} that the overall scaling degree in  $v$ should be one.  Third, we have  observed that it is sufficient to restrict  the power of $s_{ij}$ in a particular term in the  ansatz to be the same as the total number of  massive propagators in that term. Hence a general term in the ansatz  for the  pre-numerator $ \npre^{(\text{ans})}(234\cdots n-1,v)$ has  the form  
 \begin{align}
	&\Big(\prod^{d}_{h=1}{1\over v\Cdot p_{i^{(h)}_1\ldots i^{(h)}_{r_h}}}\Big) \Big(\prod^{d}_{h=1}s_{i^{(h)}j^{(h)}}\Big) \Big(\prod^{d}_{h=1}v\Cdot a_{i^{(h)}}\Big)\Big(\prod_{h=1}^{d^\prime}\varepsilon_{i^{(h)}}\Cdot a_{j^{(h)}}\Big)J_{a_2},
 \end{align}
 where $d\in [1,n-3]$, and  $a$ denotes  either $\varepsilon$ or $p$. 
 Finally, the ansatz is further constrained by   power counting  and multi-linearity with respect to external polarisation vectors (which fixes $d^\prime$). 
  
 An important comment is in order here.  The terms without any denominator  are fixed by the MHV sector in the pure Yang-Mills theory%
 \footnote{This is further discussed in Section \ref{sec:6}.} 
  with $\epsilon_1$  replaced by $m v$.  
 In fact, we have determined the MHV fusion rules, and hence the  Yang-Mills pre-numerator for this sector, which has the following form: 
    \begin{align}\label{eq:NBCJSP}
    \begin{split}
 &\npre^{(0)}_n(234\cdots n-1, v)=2^{n-3}\Bigg(\Big(\,\prod_{j=3}^{n-1} \varepsilon_{j}\Cdot p_{2\ldots j}\Big)J_{\varepsilon_2}\\
 &-\frac{1}{2} \sum_{h>\ell=2 }^{n-1}(-1)^{\ell}\Big(\,\prod_{j=\ell +1, j\neq h}^{n-1} \varepsilon_{j}\Cdot p_{2\ldots j}\Big)\varepsilon_{2}\Cdot H_{3}\Cdot H_{4}\cdots H_{\ell-1}\Cdot p_{1\ldots\ell}J_{ \varepsilon_\ell\otimes \varepsilon_h\otimes p_2}\Bigg)\,,
\end{split} 
\end{align}
where $H_{i}^{\mu\nu}=p_{2\ldots i}^{\mu}\varepsilon_{i}^{\nu}\, -\, \eta^{\mu\nu}\varepsilon_{i}\Cdot p_{2\ldots i}$. Furthermore, when  $\ell=3$ in the sum in the second line of \eqref{eq:NBCJSP}, the product of $H$  factors  should be replaced by $\eta_{\mu \nu}$, while when $\ell=2$, $\varepsilon_{2}\Cdot H_{3}\Cdot H_{4}\cdots H_{\ell-1}\Cdot p_{1\ldots\ell}$ is replaced by 1.

We will find it convenient to write the ansatz for the $n$-particle pre-numerator as the sum of the purely polynomial MHV part and a yet to be determined remainder containing massive propagators: 
 \begin{align}
 \label{ansatz}
 	\npre_n(234\cdots n-1,v):=\npre^{(0)}_n(234\cdots n-1,v)\, + \, \npre^{(\text{ans})}_n(234\cdots n-1,v)\, .
 \end{align}
In the next section we  will apply this general procedure to construct BCJ numerators in explicit examples.

 \section{Applications of the new double copy}
 \label{sec:5} 
 
 We now give examples of the construction of gauge-invariant BCJ numerators up to  six-point  amplitudes.   
We remind the reader that  in what follows $p_1 = m v \simeq -p_n$ always denote the momenta of the hard particles.

\subsection{Three-point numerator}
At three points, the  pre-numerator is simply
  \begin{align}
\npre_{3}(2, v) :=\begin{tikzpicture}[baseline={([yshift=-0.8ex]current bounding box.center)}]\tikzstyle{every node}=[font=\small]    
   \begin{feynman}
    \vertex (a)[myblob]{};
     \vertex [above=1.0cm of a](j1){$2$};
   	 \diagram*{(a) -- [thick] (j1)};
    \end{feynman}  
  \end{tikzpicture}=J_{\varepsilon_2} = m v \Cdot \varepsilon_2\, , 
 \end{align}
 and the amplitudes with a gluon or a  graviton are 
   \begin{align}
 	A^{\rm YM-M}_3(123)=\npre_{3}(2, v) \, , && A^{\rm GR-M}_3(123)=\big[\npre_{3}(2, v)\big]^2 \, .
 \end{align}
\subsection{Four-point numerator}
At four points, we know from \eqref{strconst4} that the pre-numerator is 
\begin{align}
 	\npre_4(23,v):=\begin{tikzpicture}[baseline={([yshift=-0.8ex]current bounding box.center)}]\tikzstyle{every node}=[font=\small]    
   \begin{feynman}
    \vertex (a)[myblob]{};
     \vertex [ left=.8cm of a](b);
    \vertex [above=.6cm of b](j1){$2$};
    \vertex [right=.8cm of j1](j2){};
    \vertex [right=.8cm of j2](j3){$3$};
    \vertex [right=0.5cm of j2](jnote){};
   	 \diagram*{(a) -- [thick] (j1),(a)--[thick](j3)};
    \end{feynman}  
  \end{tikzpicture}=2\Big(\frac{s_{23}v\Cdot \varepsilon _3}{4v\Cdot p_2}J_{ \varepsilon _2} -{1\over 2} J_{\varepsilon _2\otimes \varepsilon _3\otimes p_2}+\varepsilon _3\Cdot p_2 J_{ \varepsilon _2}\Big).
\end{align}
 Then the amplitudes with two gluons or two gravitons are  
 \begin{align}
 \label{4ptYMGR}
 A_4^{\rm YM-M}={\npre_4([2,3],v)\over s_{23}}, &&
 	A_4^{\rm GR-M}={\big[\npre_4([2,3],v)\big]^2\over s_{23}},
 \end{align} 
 where 
 \begin{align}
 \label{ginv4}
	\npre_4([2,3],v)&:=\begin{tikzpicture}[baseline={([yshift=-0.8ex]current bounding box.center)}]\tikzstyle{every node}=[font=\small]    
   \begin{feynman}
    \vertex (a)[myblob]{};
     \vertex [above=0.5cm of a](b)[dot]{};
     \vertex [left=0.8cm of b](c);
    \vertex [above=.5cm of c](j1){$2$};
    \vertex [right=0.8cm of j1](j2){};
    \vertex [right=0.8cm of j2](j3){$3$};
   	 \diagram*{(a) -- [thick] (b),(b) -- [thick] (j1), (b)--[thick](j3)};
    \end{feynman}  
  \end{tikzpicture}:=\mathbb L(2,3)\circ \npre_4(23,v)=2m\Big(\frac{v\Cdot F_2\Cdot F_3\Cdot v}{v\Cdot p_3}\Big).
 \end{align}
Here 
$F_i^{\mu\nu}=p_i^{\mu}\varepsilon_i^{\nu}-\varepsilon_i^{\mu}p_i^{\nu}$, and we have  used the fact that 
$J_a=m v\Cdot a$ for a vector current. Note that terms containing  tensor currents always  cancel out among themselves  after using the  Clifford algebra. For instance,  the current  $J_{\varepsilon _2\otimes \varepsilon _3\otimes p_2}$ appears in the form of a commutator, which we can recast as
\begin{align}
\begin{split}
\label{JJ}
J_{\varepsilon _2\otimes \varepsilon _3\otimes p_2}-J_{\varepsilon _3\otimes \varepsilon _2\otimes p_3}& = 
2 \varepsilon_2 \Cdot \varepsilon_3 J_{ p_2} - J_{\varepsilon _3\otimes \varepsilon _2\otimes p_{23}} = 
2 \varepsilon_2 \Cdot \varepsilon_3 J_{ p_2} - J_{\varepsilon _3\otimes \varepsilon _2\otimes p_{123}} + 
J_{\varepsilon _3\otimes \varepsilon _2\otimes p_{1}} \\  &=
2\varepsilon _2\Cdot \varepsilon _3J_{p_2}+
2 \varepsilon_2\Cdot p_1 J_{\varepsilon_3} -  2 p_1 \Cdot \varepsilon_3 J_{\varepsilon_2} - 
J_{\varepsilon _3\otimes \varepsilon _2\otimes p_{123}} 
  + J_{p_1 \otimes \varepsilon _3\otimes \varepsilon_2}
  \\ &=
2\varepsilon _2\Cdot \varepsilon _3J_{p_2}+
2 \varepsilon_2\Cdot p_1 J_{\varepsilon_3} -  2 p_1 \Cdot \varepsilon_3 J_{\varepsilon_2} 
+ m J_{\varepsilon _3\otimes \varepsilon _2} 
  -  m J_{ \varepsilon _3\otimes \varepsilon_2}
 \\
&
= 
2\varepsilon _2\Cdot \varepsilon _3J_{p_2}+
2 \varepsilon_2\Cdot p_1 J_{\varepsilon_3} -  2 p_1 \Cdot \varepsilon_3 J_{\varepsilon_2}
\ , 
\end{split}
\end{align}	
where we  have repeatedly used  \eqref{eq:CliffordAl} and  the on-shell conditions \eqref{eq:onshellJ}. 
Note the absence of tensor currents in the last line of \eqref{JJ}. We also observe that \eqref{ginv4} is expressed in terms of gauge-invariant quantities only.

\subsection{Five-point numerator} 
From five points on, we need to use the method described in Section \ref{sec:Ansatz} to determine the BCJ numerators. According to \eqref{eq:NBCJSP},  the part of the pre-numerator without linear propagators, that is $\npre^{(0)}_5(234,v)$, is given by
\begin{align}
	\npre^{(0)}_5(234,v)&= 4\Bigg(p_2\Cdot \varepsilon _3 p_{23}\Cdot \varepsilon _4 J_{\varepsilon _2}\\
	&	-\frac{1}{2} p_{23}\Cdot \varepsilon _4 J_{\varepsilon _2\otimes \varepsilon _3\otimes p_2}-\frac{1}{2} p_{23}\Cdot \varepsilon _3 J_{\varepsilon _2\otimes \varepsilon _4\otimes p_2}+\frac{1}{2} p_{23}\Cdot \varepsilon _2 J_{\varepsilon _3\otimes \varepsilon _4\otimes p_2}\Bigg).\nn
\end{align}
We now write  the ansatz for the remaining  terms in the pre-numerator, as per \eqref{ansatz}. We divide these terms into three sectors of the form
\begin{align}
	v\Cdot\varepsilon \,v\Cdot\varepsilon \, , && \varepsilon\Cdot\varepsilon \,v\Cdot \varepsilon \, , && v\Cdot\varepsilon\,  v\Cdot\varepsilon \,p\Cdot\varepsilon\, .
\end{align}
The most general ansatz  made of terms of this form contains 42 parameters, denoted by $x_{i,j}$ below: 
\begin{align}
	\npre^{(\rm ans_1)}_5(234,v)&= \Big(\frac{x_{1,1}s_{23}^2  v\Cdot \epsilon _3 v\Cdot \epsilon _4}{v\Cdot p_2 v\Cdot p_{23}}+\frac{x_{1,2}s_{24} s_{23}  v\Cdot \epsilon _3 v\Cdot \epsilon _4}{v\Cdot p_2 v\Cdot p_{23}}+
\frac{x_{1,3}s_{24}^2  v\Cdot \epsilon _3 v\Cdot \epsilon _4}{v\Cdot p_2 v\Cdot p_{23}}	\nn\\
	&+ \frac{x_{1,4}s_{34} s_{23}  v\Cdot \epsilon _3 v\Cdot \epsilon _4}{v\Cdot p_2 v\Cdot p_{23}} +\frac{x_{1,5}s_{24} s_{34}  v\Cdot \epsilon _3 v\Cdot \epsilon _4}{v\Cdot p_2 v\Cdot p_{23}}+\frac{x_{1,6}s_{34}^2  v\Cdot \epsilon _3 v\Cdot \epsilon _4}{v\Cdot p_2 v\Cdot p_{23}}\Big)J_{\epsilon _2},\nn\\
\npre^{(\rm ans_2)}_5(234,v)&= \Big(\frac{x_{2,1}s_{23}  \epsilon _3\Cdot \epsilon _4 v\Cdot p_3}{v\Cdot p_2}+\text{11 terms}\Big)J_{\epsilon _2}\nn\\
\npre^{(\rm ans_3)}_5(234,v)&=\Big(\frac{x_{3,1} s_{23}  \epsilon _4\Cdot p_3 v\Cdot \epsilon _3}{v\Cdot p_2}+\text{23 terms}\Big)J_{\epsilon _2}.
\end{align}
We solve for the parameters according to \eqref{eq:newDC},  and  arrive at a pre-numerator with 11 undetermined  parameters: 
\begin{align}
 	\npre_5(234,v)&:=\begin{tikzpicture}[baseline={([yshift=-0.8ex]current bounding box.center)}]\tikzstyle{every node}=[font=\small]    
   \begin{feynman}
    \vertex (a)[myblob]{};
     \vertex [ left=.8cm of a](b);
    \vertex [above=.6cm of b](j1){$2$};
    \vertex [right=.8cm of j1](j2){$3$};
    \vertex [right=.8cm of j2](j3){$4$};
    \vertex [right=0.5cm of j2](jnote){};
   	 \diagram*{(a) -- [thick] (j1),(a)--[thick](j2),(a)--[thick](j3)};
    \end{feynman}  
  \end{tikzpicture}\\
	&=4\Big(\frac{s_{23} \epsilon _4\Cdot p_3 v\Cdot \epsilon _3}{2v\Cdot p_2}+\frac{s_{34} \epsilon _3\Cdot p_2 v\Cdot \epsilon _4}{2v\Cdot p_{23}}+\frac{s_{23}  \epsilon _3\Cdot \epsilon _4 v\Cdot p_4}{2v\Cdot p_2}+\frac{s_{23} s_{34}  v\Cdot \epsilon _3 v\Cdot \epsilon _4}{8 v\Cdot p_2 v\Cdot p_{23}} \Big)J_{\epsilon _2}\nn\\
	&+ \npre^{\rm (0)}_5(234,v)+x_{1,1}\left(\frac{s_{23}^2 v\Cdot \epsilon _3 v\Cdot \epsilon _4}{v\Cdot p_2 v\Cdot p_{23}}-\frac{s_{34}^2 v\Cdot \epsilon _3 v\Cdot \epsilon _4}{v\Cdot p_2 v\Cdot p_{23}}\right)J_{\epsilon _2} +\cdots,\nn 
\end{align}
where we have indicated for brevity only one of the terms with a free parameter. 
Remarkably, we find that the terms with undetermined parameters   vanish under the action of $\mathbb L(2,3,4)$:
\begin{align}
\label{dropout}
	\mathbb L(2,3,4) \circ \Big[x_{1,1}\left(\frac{s_{23}^2 v\Cdot \epsilon _3 v\Cdot \epsilon _4}{v\Cdot p_2 v\Cdot p_{23}}-\frac{s_{34}^2 v\Cdot \epsilon _3 v\Cdot \epsilon _4}{v\Cdot p_2 v\Cdot p_{23}}\right)J_{\epsilon _2} +\cdots\Big]=  0 \ . 
\end{align}
As a consequence, and in contradistinction with the expectation from pure Yang-Mills amplitudes, we arrive at a unique solution for the BCJ numerator.

Summarising, the BCJ numerator for the left-nested commutator is 
\begin{align}
&&\npre_5([[2,3],4],v):=\begin{tikzpicture}[baseline={([yshift=-0.8ex]current bounding box.center)}]\tikzstyle{every node}=[font=\small]    
   \begin{feynman}
    \vertex (a)[myblob]{};
     \vertex [above=0.5cm of a](b)[dot]{};
     \vertex [left=1.0cm of b](c);
     \vertex [left=0.5cm of b](c23);
     \vertex [above=0.35cm of c23](v23)[dot]{};
    \vertex [above=.6cm of c](j1){$2$};
    \vertex [right=1.0cm of j1](j2){$3$};
    \vertex [right=1.0cm of j2](j3){$4$};
   	 \diagram*{(a) -- [thick] (b),(b) -- [thick] (j1),(v23) -- [thick] (j2),(b)--[thick](j3)};
    \end{feynman}  
  \end{tikzpicture}:= \mathbb L(2,3,4) \circ \npre_5(234,v)
\ .
\end{align}
Importantly, we find that this  BCJ numerator can be  rewritten in a manifestly gauge-invariant form as
\begin{align}\label{eq:FiveBCJNum}
	\npre_5([[2,3],4],v)
	&= \mathbb{L}(2,3,4)\circ \Big[4m {v\Cdot F_2\Cdot F_3\Cdot V_3\Cdot F_4\Cdot v\over v\Cdot p_3 v\Cdot p_{4}}\Big]\, , 
\end{align}
where $V_i^{\mu\nu}=v^{\mu}p_i^\nu$. 
Then the colour-ordered amplitude in  the novel colour-kinematics duality form is 
\begin{align}
	A^{\rm YM-M}_5(12345)={\npre_5([[2,3],4],v)\over s_{234} s_{23}} +{\npre_5([2,[3,4]],v)\over s_{234} s_{34}}\, .
\end{align}
Note that these two terms are related by exchanging $2\leftrightarrow 4$. By the double copy,  the gravity amplitude is obtained immediately as
\begin{align}
\begin{split}
\label{A5new}
A^{\rm GR-M}_5(12345)&=
{\big[\npre_5([[2,3],4],v)\big]^2 \over s_{234} s_{23}} + {\big[\npre_5([[2,4],3],v)\big]^2 \over s_{234} s_{24}}+{\big[\npre_5([[3,4],2],v)\big]^2 \over s_{234} s_{34}}
\, .
\end{split}
\end{align}
We have checked that  \eqref{A5new} agrees with  the $\mathcal{O}(m^2)$ term of   \eqref{eq:A5GR}. 
It is useful to pause and contrast these two expressions to appreciate the simplicity of our approach. First,  
  \eqref{A5new} is much more compact  than  \eqref{eq:A5GR}; furthermore, expanding  \eqref{eq:A5GR} to $\mathcal{O}(m^2)$ gives rise to a very large number of terms lacking any particular structure. The compactness, and manifest gauge invariance and locality of our HEFT amplitudes are of considerable advantage  when these expressions are fed into unitarity cuts.

We also comment that the numerators we have constructed automatically satisfy the Jacobi relations. For instance,   by definition we have 
\begin{align}
	\npre_5([[2,3],4],v)&=\npre_5(234,v)-\npre_5(324,v)-\npre_5(423,v)+\npre_5(432,v)\nn\\
	\npre_5([[2,4],3],v)&=\npre_5(243,v)-\npre_5(423,v)-\npre_5(324,v)+\npre_5(342,v)\nn\\
	\npre_5([2,[3,4]],v)&=\npre_5(234,v)-\npre_5(243,v)-\npre_5(342,v)+\npre_5(432,v)\, , 
\end{align}
which makes the Jacobi relation manifest, 
\begin{align}
\npre_5([2,[3,4]],v)=	\npre_5([[2,3],4],v)-\npre_5([[2,4],3],v)\, .
\end{align}
Finally, note that one can construct   $\npre_5([[2,4],3],v)$ from  $\npre_5([[2,3],4],v)$  by simply swapping labels  
3 and 4.

\subsection{Six-point numerator} 
Similarly to what was done at five points, using again \eqref{eq:NBCJSP} we find that, at six points, the polynomial part of the pre-numerator, $\npre^{(0)}_6(2345,v)$, has the form 
\begin{align}
	&\npre^{(0)}_6(2345,v)=8\Bigg(J_{\varepsilon _2} p_2\Cdot \varepsilon _3 p_{23}\Cdot \varepsilon _4 p_{234}\Cdot \varepsilon _5 \\
	&-\frac{1}{2} p_{23}\Cdot \varepsilon _4 p_{234}\Cdot \varepsilon _5 J_{\varepsilon _2\otimes \varepsilon _3\otimes p_2}-\frac{1}{2} p_2\Cdot \varepsilon _3 p_{234}\Cdot \varepsilon _5 J_{\varepsilon _2\otimes \varepsilon _4\otimes p_2}-\frac{1}{2} p_2\Cdot \varepsilon _3 p_{23}\Cdot \varepsilon _4 J_{\varepsilon _2\otimes \varepsilon _5\otimes p_2}	 \nn\\
	&+\frac{1}{2} p_3\Cdot \varepsilon _2 p_{234}\Cdot \varepsilon _5 J_{\varepsilon _3\otimes \varepsilon _4\otimes p_2}+\frac{1}{2} p_3\Cdot \varepsilon _2 p_{23}\Cdot \varepsilon _4 J_{\varepsilon _3\otimes \varepsilon _5\otimes p_2}+\frac{1}{2} \left(\varepsilon _2\Cdot p_4 \varepsilon _3\Cdot p_2-\varepsilon _2\Cdot p_3 \varepsilon _3\Cdot p_4\right) J_{\varepsilon _4\otimes \varepsilon _5\otimes p_2}\Bigg).\nn
\end{align}
The most general  ansatz for the remainder of the pre-numerator can be decomposed into six sectors.  The number of free parameters and    physical constraints arising from \eqref{eq:newDC} are shown in the table below:
\begin{align}
\begin{array}{|c||c|c|c|c|c|c|c|}
  \hline
	\text{sector} & v\Cdot\varepsilon\, v\Cdot\varepsilon \,v\Cdot\varepsilon \,v\Cdot\varepsilon 
	& v\Cdot\varepsilon\, v\Cdot\varepsilon \,v\Cdot\varepsilon \,p\Cdot \varepsilon 
	& v\Cdot\varepsilon\, v\Cdot\varepsilon \,\varepsilon\Cdot\varepsilon 
	&  v\Cdot\varepsilon\, v\Cdot\varepsilon \, p\Cdot \varepsilon \,p\Cdot \varepsilon 
	& v\Cdot\varepsilon\, \varepsilon\Cdot\varepsilon \,p\Cdot \varepsilon 
	&  \varepsilon\Cdot\varepsilon \,\varepsilon\Cdot\varepsilon \\
	 \hline  \hline
	 \text{\# of parameters} &
	280 &  1134 &  945 &  1128  & 1134  & 378 \\
	 \hline
	\text{\# of constraints} & 176 &  601  & 438 &  570 &  810  & 108\\
	 \hline
\end{array}
\end{align}
Thus  we arrive at a pre-numerator with 2296 free parameters:
\begin{align}
& 	\npre_6(2345,v):=\begin{tikzpicture}[baseline={([yshift=-0.8ex]current bounding box.center)}]\tikzstyle{every node}=[font=\small]    
   \begin{feynman}
    \vertex (a)[myblob]{};
     \vertex [ left=1.2cm of a](b);
    \vertex [above=.6cm of b](j1){$2$};
    \vertex [right=.8cm of j1](j2){$3$};
    \vertex [right=.8cm of j2](j3){$4$};
    \vertex [right=.8cm of j3](j4){$5$};
   	 \diagram*{(a) -- [thick] (j1),(a)--[thick](j2),(a)--[thick](j3),(a)--[thick](j4)};
    \end{feynman}  
  \end{tikzpicture}=8\Bigg({s_{34}v\Cdot \varepsilon _3 v\Cdot \varepsilon _4 v\Cdot \varepsilon _5 \over 8v\Cdot p_2 v\Cdot p_{23}}\Big(\frac{s_{23} s_{45} }{2  v\Cdot p_{234}}+\frac{s_{24} s_{45} }{3  v\Cdot p_{234}}-\frac{s_{23}  s_{35}}{3 v\Cdot p_4}\Big)J_{\varepsilon _2}\nn\\
	&+\Big(\frac{s_{23} s_{34}  \varepsilon _5\Cdot p_4 v\Cdot \varepsilon _3 v\Cdot \varepsilon _4}{4v\Cdot p_2 v\Cdot p_{23}}+\frac{s_{45} s_{34}  \varepsilon _3\Cdot p_2 v\Cdot \varepsilon _4 v\Cdot \varepsilon _5}{4v\Cdot p_{23} v\Cdot p_{234}}-\frac{s_{23} s_{34}  \varepsilon _5\Cdot p_3 v\Cdot \varepsilon _3 v\Cdot \varepsilon _4}{8 v\Cdot p_2 v\Cdot p_4}\nn\\
	&~~~~-\frac{s_{35} s_{34}  \varepsilon _3\Cdot p_2 v\Cdot \varepsilon _4 v\Cdot \varepsilon _5}{4v\Cdot p_4 v\Cdot p_{234}}-\frac{s_{23} s_{45}  \varepsilon _3\Cdot p_4 v\Cdot \varepsilon _4 v\Cdot \varepsilon _5}{4v\Cdot p_2 v\Cdot p_{234}}-\frac{s_{24} s_{45} \varepsilon _3\Cdot p_4 v\Cdot \varepsilon _4 v\Cdot \varepsilon _5}{8 v\Cdot p_2 v\Cdot p_{234}}\Big)J_{\varepsilon _2} \nn\\
	&+\Big(\frac{s_{23} s_{34}  \varepsilon _4\Cdot \varepsilon _5 v\Cdot p_5 v\Cdot \varepsilon _3}{4v\Cdot p_2 v\Cdot p_{23}}-\frac{s_{23} s_{34}  \varepsilon _3\Cdot \varepsilon _5 v\Cdot p_5 v\Cdot \varepsilon _4}{8 v\Cdot p_2 v\Cdot p_4}+\frac{s_{23} s_{35}  \varepsilon _3\Cdot \varepsilon _4 v\Cdot p_4 v\Cdot \varepsilon _5}{8 v\Cdot p_{23} v\Cdot p_{234}}\nn\\
	&~~~~+{s_{23} s_{45}  \varepsilon _3\Cdot \varepsilon _4 v\Cdot \varepsilon _5\over 8v\Cdot p_{234}}(\frac{ v\Cdot p_4 }{ v\Cdot p_2 }+\frac{ v\Cdot p_{45} }{v\Cdot p_{23} })+\frac{s_{24} s_{45}\varepsilon _3\Cdot \varepsilon _4 v\Cdot p_4 v\Cdot \varepsilon _5}{8  v\Cdot p_{234}}(\frac{1}{ v\Cdot p_2 }-\frac{1}{v\Cdot p_{23} })\Big)J_{\varepsilon _2} \nn\\
	&+ \Big(\frac{(s_{23}-s_{45}) \varepsilon _3\Cdot p_4 \varepsilon _5\Cdot p_2 v\Cdot \varepsilon _4}{2v\Cdot p_{34}}+s_{23} \varepsilon _4\Cdot p_3 v\Cdot \varepsilon _3(\frac{ \varepsilon _5\Cdot p_{34} }{2v\Cdot p_2}-\frac{ \varepsilon _5\Cdot p_2 }{2v\Cdot p_{34}})+\frac{s_{25} \varepsilon _3\Cdot p_2 \varepsilon _5\Cdot p_4 v\Cdot \varepsilon _4}{2v\Cdot p_{23}}\nn\\
	&~~~~+\frac{(s_{35} +s_{45})\varepsilon _3\Cdot p_2 \varepsilon _4\Cdot p_{23} v\Cdot \varepsilon _5}{2v\Cdot p_{234}}-\frac{s_{34} \varepsilon _3\Cdot p_2 \varepsilon _5\Cdot p_{23} v\Cdot \varepsilon _4}{2v\Cdot p_4}-\frac{(s_{25}+s_{34}) \varepsilon _3\Cdot p_2 \varepsilon _4\Cdot p_5 v\Cdot \varepsilon _5}{4 v\Cdot p_{23}}\Big)J_{\varepsilon _2}\nn\\
	& +\Big(\frac{s_{23} \varepsilon _3\Cdot \varepsilon _5 v\Cdot p_5 \varepsilon _4\Cdot p_2}{2v\Cdot p_{24}}+\frac{s_{23} \varepsilon _3\Cdot \varepsilon _5 v\Cdot p_5 \varepsilon _4\Cdot p_3}{2v\Cdot p_2}+\frac{s_{23} \varepsilon _3\Cdot \varepsilon _4 v\Cdot p_4 \varepsilon _5\Cdot p_3}{2v\Cdot p_2}+\frac{s_{23} \varepsilon _3\Cdot \varepsilon _4 v\Cdot p_{45} \varepsilon _5\Cdot p_4}{2v\Cdot p_2}\nn\\
	&~~~~-\frac{s_{23} \varepsilon _4\Cdot \varepsilon _5 v\Cdot p_5 \varepsilon _3\Cdot p_4}{2v\Cdot p_2}-\frac{s_{23} \varepsilon _3\Cdot \varepsilon _4 v\Cdot p_4 \varepsilon _5\Cdot p_2}{2v\Cdot p_{34}}+\frac{(s_{24}+s_{34}) \varepsilon _4\Cdot \varepsilon _5 v\Cdot p_5 \varepsilon _3\Cdot p_2}{2v\Cdot p_{23}}\Big)J_{\varepsilon _2}  \nn\\
	&+\Big(\frac{s_{25} \varepsilon _2\Cdot \varepsilon _3 \varepsilon _4\Cdot \varepsilon _5 v\Cdot p_4}{4 v\Cdot p_{23}}+\frac{s_{35} \varepsilon _2\Cdot \varepsilon _3 \varepsilon _4\Cdot \varepsilon _5 v\Cdot p_5}{4 v\Cdot p_{234}}+\frac{s_{45} \varepsilon _2\Cdot \varepsilon _4 \varepsilon _3\Cdot \varepsilon _5 v\Cdot p_5}{4 v\Cdot p_{234}}\nn\\
	&~~~~-s_{45} \varepsilon _2\Cdot \varepsilon _5 \varepsilon _3\Cdot \varepsilon _4(\frac{ v\Cdot p_4}{2v\Cdot p_{34}}+\frac{ v\Cdot p_5}{4 v\Cdot p_{234}})\Big)J_{p_2}\Bigg)+\npre^{\rm (0)}_6(2345,v)+\cdots , 
\end{align}
where we omit all the terms with free parameters. Then the BCJ numerator from the left-nested commutator is obtained by acting with the operator $\mathbb L(2,3,4,5)$:
\begin{align}
	\npre_6([[[2,3],4],5],v):=\begin{tikzpicture}[baseline={([yshift=-0.8ex]current bounding box.center)}]\tikzstyle{every node}=[font=\small]    
   \begin{feynman}
    \vertex (a)[myblob]{};
     \vertex [above=0.5cm of a](b)[dot]{};
     \vertex [left=1.5cm of b](c);
     \vertex [left=1.0cm of b](c23);
      \vertex [left=0.5cm of b](c234);
     \vertex [above=0.49cm of c23](v23)[dot]{};
       \vertex [above=0.24cm of c234](v234)[dot]{};
    \vertex [above=.6cm of c](j1){$2$};
    \vertex [right=1.0cm of j1](j2){$3$};
    \vertex [right=1.0cm of j2](j3){$4$};
     \vertex [right=1.0cm of j3](j4){$5$};
   	 \diagram*{(a) -- [thick] (b),(b) -- [thick] (j1),(v23) -- [thick] (j2),(v234) -- [thick] (j3),(b)--[thick](j4)};
    \end{feynman}  
  \end{tikzpicture}:=\mathbb L(2,3,4,5) \circ  \npre_6(2345,v)\, .
\end{align}
The  BCJ numerator thus obtained  does not contain any free parameter and is unique:
\begin{align}\label{eq:SixBCJNum}
	&\npre_6([[[2,3],4],5],v)
	= 8m\,\mathbb{L}(2, 3, 4, 5)\circ\Big[\\
	&\frac{v\Cdot F_2\Cdot F_3\Cdot V_3\Cdot F_4\Cdot F_5\Cdot v}{2 v\Cdot p_3 v\Cdot p_{45}}+\frac{v\Cdot F_2\Cdot F_3\Cdot V_3\Cdot F_4\Cdot V_4\Cdot F_5\Cdot v}{2 v\Cdot p_3 v\Cdot p_5 v\Cdot p_{45}}+\frac{v\Cdot F_4\Cdot F_5\Cdot V_3\Cdot F_2\Cdot V_4\Cdot F_3\Cdot v}{2 v\Cdot p_3 v\Cdot p_4 v\Cdot p_{345}}\nn\\
	&-\frac{v\Cdot F_3\Cdot F_5\Cdot V_3\Cdot F_2\Cdot V_3\Cdot F_4\Cdot v}{2 v\Cdot p_3 v\Cdot p_4 v\Cdot p_{345}}-\frac{v\Cdot F_3\Cdot F_4\Cdot V_3\Cdot F_2\Cdot V_3\Cdot F_5\Cdot v}{2 v\Cdot p_3 v\Cdot p_5 v\Cdot p_{345}}-\frac{v\Cdot p_{45} v\Cdot F_3\Cdot F_4\Cdot V_3\Cdot F_2\Cdot V_4\Cdot F_5\Cdot v}{2 v\Cdot p_3 v\Cdot p_4 v\Cdot p_5 v\Cdot p_{345}}
	\Big]. \nn
\end{align}
The colour-ordered amplitude in  the novel colour-kinematics duality form is then  
\begin{align}
\begin{split}
	A^{\rm YM-M}_6(123456)&={\npre_6([[[2,3],4],5],v)\over s_{2345}s_{234} s_{23}} +{\npre_6([2,[3,[4,5]]],v)\over s_{2345}s_{345} s_{45}}+{\npre_6([[2,3],[4,5]],v)\over s_{2345}s_{23} s_{45}}\\
	&+{\npre_6([[2,[3,4]],5],v)\over s_{2345}s_{234} s_{34}}+{\npre_6([2,[[3,4],5]],v)\over s_{2345}s_{34} s_{345}}\, .
\end{split}
\end{align}
The gravity amplitude is constructed similarly from the double copy:
\begin{align}
\label{BCJ-6}
\begin{split}
	\hspace{-0.5cm} 
	A_6^{\rm GR-M}(123456)&={[\npre_6({[[2,3],[4,5]]},v)]^2\over s_{23}s_{45}s_{2345}}+{[\npre_6({[2,[3,[4,5]]]},v)]^2\over s_{345}s_{45}s_{2345}}+{[\npre_6({[[2,4],[3,5]]},v)]^2\over s_{24}s_{35}s_{2345}}\\
	&+{[\npre_6({[2,[4,[3,5]]]},v)]^2\over s_{345}s_{35}s_{2345}}+{[\npre_6({[[2,5],[3,4]]},v)]^2\over s_{25}s_{34}s_{2345}}+{[\npre_6({[2,[5,[3,4]]]},v)]^2\over s_{345}s_{34}s_{2345}}\\
	&+{[\npre_6({[3,[2,[4,5]]]},v)]^2\over s_{245}s_{45}s_{2345}}+{[\npre_6({[3,[4,[2,5]]]},v)]^2\over s_{245}s_{25}s_{2345}}+{[\npre_6({[3,[5,[2,4]]]},v)]^2\over s_{24}s_{245}s_{2345}}\\
	&+{[\npre_6({[4,[2,[3,5]]]},v)]^2\over s_{235}s_{35}s_{2345}}+{[\npre_6({[4,[3,[2,5]]]},v)]^2\over s_{235}s_{25}s_{2345}}+{[\npre_6({[4,[5,[2,3]]]},v)]^2\over s_{23}s_{235}s_{2345}}\\
	&+{[\npre_6({[5,[2,[3,4]]]},v)]^2\over s_{234}s_{34}s_{2345}}+{[\npre_6({[5,[3,[2,4]]]},v)]^2\over s_{24}s_{234}s_{2345}}+{[\npre_6({[5,[4,[2,3]]]},v)]^2\over s_{23}s_{234}s_{2345}}\\ 
	 & =  
	 {[\npre_6({[[2,3],[4,5]]},v)]^2\over 8s_{23}s_{45}s_{2345}}+{[\npre_6({[2,[3,[4,5]]]},v)]^2\over 2s_{345}s_{45}s_{2345}}
	 +\text{permutations of 2,3,4,5}.
\end{split}
\end{align}
Note that gauge invariance, crossing symmetry and locality for the massless particles in the BCJ numerator \eqref{BCJ-6} are manifest.

\section{Local BCJ numerators for pure Yang-Mills theory}
\label{sec:6}

The numerators we have constructed in the HEFT are closely related to the local BCJ numerators for  pure Yang-Mills theory by the following equation
\begin{align}
	N_{n-1}^{\rm YM}(234\cdots (n-1) 1)=\npre_n([\cdots [[2,3],4],\cdots,n-1],v)|_{v\rightarrow \varepsilon_1,p_{234\cdots n-1}^2=0}
	\, .
\end{align}
We have  checked this relation at five and six points, and we will now illustrate it in detail in the five-point case.

At five points, the main observation is to  impose  the on-shell condition for the leg with momentum $p_{234}$. Doing so,   we get 
\begin{align}
	N^{\rm YM}_4(2341)=\npre_5([[2,3],4],v)|_{mv\rightarrow \varepsilon_1,p_{234}^2=0}.
\end{align}
We expand   $\npre_5([[2,3],4],v)$ in \eqref{eq:FiveBCJNum} and get 
\begin{align}\label{lunga}
	N_4^{\rm YM}(2341)&=4\Big(\varepsilon _1\Cdot \varepsilon _4 \varepsilon _2\Cdot p_3 \varepsilon _3\Cdot p_4-\varepsilon _1\Cdot \varepsilon _4 \varepsilon _2\Cdot p_4 \varepsilon _3\Cdot p_2+\varepsilon _2\Cdot \varepsilon _4 \varepsilon _1\Cdot p_4 \varepsilon _3\Cdot p_2+\varepsilon _1\Cdot \varepsilon _2 \varepsilon _4\Cdot p_{23} \varepsilon _3\Cdot p_2\nn\\
	&-\varepsilon _3\Cdot \varepsilon _4 \varepsilon _1\Cdot p_4 \varepsilon _2\Cdot p_3-\varepsilon _1\Cdot \varepsilon _3 \varepsilon _2\Cdot p_3 \varepsilon _4\Cdot p_{23}+\varepsilon _2\Cdot \varepsilon _3 \varepsilon _1\Cdot p_3 \varepsilon _4\Cdot p_{23}+\varepsilon _2\Cdot \varepsilon _3 \varepsilon _1\Cdot p_4 \varepsilon _4\Cdot p_3\nn\\
	&+ p_2\Cdot p_{34} \varepsilon _1\Cdot \varepsilon _4 \varepsilon _2\Cdot \varepsilon _3-{1\over 4}s_{23} \varepsilon _1\Cdot \varepsilon _2 \varepsilon _3\Cdot \varepsilon _4+{1\over 2}s_{23}\mathcal I (3,4)\Big)
\, ,
\end{align}
where 
\begin{align}
	\mathcal I (3,4)=&-{1\over 2}\frac{ \varepsilon _1\Cdot \varepsilon _2 \varepsilon _3\Cdot \varepsilon _4 }{\varepsilon _1\Cdot p_{34}}(\varepsilon _1\Cdot p_4-\varepsilon _1\Cdot p_3)+\frac{\varepsilon _1\Cdot \varepsilon _2 \varepsilon _1\Cdot \varepsilon _3 \varepsilon _1\Cdot \varepsilon _4}{ \varepsilon _1\Cdot p_{34}}\Big(\frac{s_{24}}{\varepsilon _1\Cdot p_3}-\frac{s_{23} }{\varepsilon _1\Cdot p_4 }\Big)\nn\\
	&+\varepsilon _1\Cdot \varepsilon _3 \varepsilon _1\Cdot \varepsilon _4\Big(\frac{  \varepsilon _2\Cdot p_4}{\varepsilon _1\Cdot p_3}-\frac{ \varepsilon _2\Cdot p_3}{\varepsilon _1\Cdot p_4}\Big)+\frac{  \varepsilon _1\Cdot \varepsilon _2 }{\varepsilon _1\Cdot p_{34}}( \varepsilon _1\Cdot \varepsilon _4 \varepsilon _3\Cdot p_4-\varepsilon _1\Cdot \varepsilon _3 \varepsilon _4\Cdot p_3)\nn\\
	&+\frac{ \varepsilon _1\Cdot \varepsilon _4 \varepsilon _2\Cdot \varepsilon _3 \varepsilon _1\Cdot p_3}{\varepsilon _1\Cdot p_4}+\frac{ \varepsilon _1\Cdot \varepsilon _2 \varepsilon _1\Cdot \varepsilon _4 \varepsilon _3\Cdot p_2}{\varepsilon _1\Cdot p_4}-\frac{\varepsilon _1\Cdot \varepsilon _3 \varepsilon _2\Cdot \varepsilon _4 \varepsilon _1\Cdot p_4}{\varepsilon _1\Cdot p_3}-\frac{ \varepsilon _1\Cdot \varepsilon _2 \varepsilon _1\Cdot \varepsilon _3 \varepsilon _4\Cdot p_2}{\varepsilon _1\Cdot p_3}.
\end{align}
As ${1\over 2}\mathbb L(3,4)\circ \mathcal I(3,4)= \mathcal I(3,4)$, then $s_{23}  \mathcal I(3,4)$ is a pure gauge term\footnote{We comment that 
 pure gauge terms are classified by the  $\mathbb L$-invariants $\mathcal{I}$  
 as discussed in~\cite{Chen:2021chy}.}  which does not contribute to the Yang-Mills amplitude. 
By removing pure gauge terms we   arrive at an expression which is manifestly local,
 \begin{align}
 \label{finalnumnum}
	N^{\rm YM}_4(2341)&=4\Big(\varepsilon _1\Cdot \varepsilon _4 \varepsilon _2\Cdot p_3 \varepsilon _3\Cdot p_4-\varepsilon _1\Cdot \varepsilon _4 \varepsilon _2\Cdot p_4 \varepsilon _3\Cdot p_2+\varepsilon _2\Cdot \varepsilon _4 \varepsilon _1\Cdot p_4 \varepsilon _3\Cdot p_2+\varepsilon _1\Cdot \varepsilon _2 \varepsilon _4\Cdot p_{23} \varepsilon _3\Cdot p_2\nn\\
	&-\varepsilon _3\Cdot \varepsilon _4 \varepsilon _1\Cdot p_4 \varepsilon _2\Cdot p_3-\varepsilon _1\Cdot \varepsilon _3 \varepsilon _2\Cdot p_3 \varepsilon _4\Cdot p_{23}+\varepsilon _2\Cdot \varepsilon _3 \varepsilon _1\Cdot p_3 \varepsilon _4\Cdot p_{23}+\varepsilon _2\Cdot \varepsilon _3 \varepsilon _1\Cdot p_4 \varepsilon _4\Cdot p_3\nn\\
	&+ p_2\Cdot p_{34} \varepsilon _1\Cdot \varepsilon _4 \varepsilon _2\Cdot \varepsilon _3-{1\over 4}s_{23} \varepsilon _1\Cdot \varepsilon _2 \varepsilon _3\Cdot \varepsilon _4\Big).
\end{align}
 Finally, we have  checked that the  numerator \eqref{finalnumnum} generates the correct amplitude. This  numerator by itself is no longer  gauge invariant.

We have also confirmed that the six-point numerator in HEFT is  
 directly related  to the five-point BCJ numerator in pure Yang-Mills theory,
\begin{align}
	N_5^{\rm YM}(23451)=\npre_6([[[2,3],4],5],v)|_{v\rightarrow \varepsilon_1,p_{2345}^2=0}\, .
\end{align}
This result is in agreement with   \cite{Chen:2019ywi}.

\section{Conclusions}
\label{sec:7}

We conclude by  summarising a number of interesting properties  of the new BCJ numerators we have constructed in the HEFT: 
\begin{itemize}
	\item[{\bf 1.}]  {\bf Gauge invariance.} We have expressed all the BCJ numerators   $\npre_n(\commut,v)$ in terms of field strengths,  hence they are gauge invariant  
	even if the on-shell   and transversality conditions  $p_i\Cdot \varepsilon_i=0$ are not imposed.
	\item[{\bf 2.}] {\bf Locality with respect to the massless particles.} Our new BCJ numerators do not contain  spurious poles. This  is particularly convenient when  constructing loop integrands via generalised unitarity, as we will demonstrate in several examples at loop level in \cite{companion2}. 
	\item[{\bf 3.}] {\bf Crossing symmetry and Jacobi relation are manifest.} As all the BCJ numerators are generated from the pre-numerator,  the crossing symmetry with respect to all the massless particles is manifest. Moreover, based on the assumption of  associativity of the fusion product,  the Jacobi relations are automatically satisfied by the BCJ numerators.
	
\item[{\bf 4.}]
	{\bf Numerators for pure Yang-Mills theory}.
Our BCJ numerators are directly related to local BCJ numerators for pure Yang-Mills theory. 

\end{itemize}
Note that usually BCJ numerators are manifestly local but not gauge invariant, while using KLT relations one arrives at expressions which are gauge invariant but non-local; here we have both locality and gauge invariance. 
We have checked the above properties up to six particles, and  we conjecture them to be valid for arbitrary number of particles. 

There are a few directions for future work. 
First, it would be desirable to find a proof of the gauge invariance of the BCJ numerators for any multiplicity. Is this a property of the HEFT or only of its leading term considered in this paper? A pressing question is to find a closed form of the pre-numerator, which would require a full understanding of the  fusion rule  for any number of particles. Finally, it would also be important to understand the subleading terms in the inverse mass expansion as well as  higher-spin effects, to ascertain how much of the structures  we have uncovered survive the expansion. 


\section*{Acknowledgements}

We would like to thank   Henrik Johansson, Rodolfo Russo, Fei Teng and  Tianheng Wang 
for interesting discussions. This work  was supported by the Science and Technology Facilities Council (STFC) Consolidated Grants ST/P000754/1 \textit{``String theory, gauge theory \& duality''} and  ST/T000686/1 \textit{``Amplitudes, strings  \& duality''}
and by the European Union's Horizon 2020 research and innovation programme under the Marie Sk\l{}odowska-Curie grant agreement No.~764850 {\it ``\href{https://sagex.org}{SAGEX}''}.
CW is supported by a Royal Society University Research Fellowship No.~UF160350.

  \newpage

\bibliographystyle{JHEP}
\bibliography{ScatEq}

\end{document}